# The Solar Corona during the Total Eclipse on 16 June 1806: Graphical Evidence of the Coronal Structure during the Dalton Minimum

Hisashi Hayakawa (1-3)*, Mathew J. Owens (4), Michael Lockwood (4), Mitsuru Sôma (5)

(1) Institute for Space-Earth Environmental Research, Nagoya University, Nagoya, 4648601, Japan
(2) Institute for Advanced Researches, Nagoya University, Nagoya, 4648601, Japan
(3) Science and Technology Facilities Council, RAL Space, Rutherford Appleton Laboratory, Harwell Campus, Didcot, OX11 0QX, UK
(4) Department of Meteorology, University of Reading, Reading RG6 6BB, UK
(5) National Astronomical Observatory of Japan, Mitaka, 1818588, Japan

* hisashi@nagoya-u.jp; hisashi.hayakawa@stfc.ac.uk

**Abstract**

Visible coronal structure, in particular the spatial evolution of coronal streamers, provides indirect information about solar magnetic activity and the underlying solar dynamo. Their apparent absence of structure observed during the total eclipses of throughout the Maunder Minimum has been interpreted as evidence of a significant change in the solar magnetic field from that during modern cycles. Eclipse observations available from the more recent Dalton Minimum may be able to provide further information, sunspot activity being between the levels seen during recent cycles and in the Maunder Minimum. Here, we show and examine two graphical records of the total solar eclipse on 1806 June 16, during the Dalton Minimum. These records show significant rays and streamers around an inner ring. The ring is estimated to be ≈ 0.44 $R_{\odot}$ in width and the streamers in excess of 11.88 $R_{\odot}$ in length. In combination with records of spicules or prominences, these eclipse records visually contrast the Dalton Minimum with the Maunder Minimum in terms of their coronal structure and support the existing discussions based on the sunspot observations. These eclipse records are broadly consistent with the solar cycle phase in the modelled open solar flux and the reconstructed slow solar wind at most latitudes.

**1. Introduction**

Variability of the solar magnetic field has been directly monitored for ≈ 4 centuries with sunspot observations as a visual manifestation of magnetic flux (Clette *et al*., 2014; Arlt and Vaquero, 2020). These observations show the regular Schwabe cycle of ≈ 11 years and two longer-term intervals with significantly suppressed solar activity: most prominently, the Maunder Minimum (hereafter MM; *c*., 1645 – 1715) and, to a somewhat lesser extent, the Dalton Minimum (hereafter, DM; *c*., 1797 – 1827) (Hathaway, 2015; Muñoz-Jaramillo and Vaquero, 2019). While a number of additional intervals with comparable solar activity have been identified over millennial time scales using proxy reconstructions with the cosmogenic isotopes (Usoskin *et al*., 2007; Inceoglu *et al*., 2015), only the MM and DM can be investigated with direct observations and measurements (Usoskin *et al*., 2015; Hayakawa *et al*., 2020).

The physical nature of these two intervals, the MM and the DM, is of great interest as grand minima are generally associated with a special state of the solar dynamo (Charbonneau, 2010). Analyses of these intervals are difficult, due to their poor observational coverage and different observational motivations relative to the modern era (Arlt and Vaquero, 2020). Nevertheless, thorough analyses on the original observations have revealed their differences in terms of their solar-cycle amplitude and length, as well as sunspot distributions and highlighted their probable difference, although the poor observational coverage still prevents definitive conclusions (Eddy, 1976; Ribes and Nesme-Ribes, 1993; Usoskin *et al*., 2015; Hayakawa *et al*., 2020) and even accommodates discussions on possibility of one solar cycle lost just before the onset of the DM (Usoskin *et al*., 2009; Karoff *et al*., 2015; Owens *et al*., 2015; Vaquero *et al*., 2016; Hayakawa *et al*., 2018).

In this regard, the solar coronal structure is of significant interest, forming a visual representation of the large-scale solar magnetic field, and with the solar coronal holes providing a visual estimate of the extent of the fast solar wind source regions. In the typical solar cycles of the modern era, the polar coronal holes reach maximum areal extent around the minima to concentrate the coronal streamers nearer the solar equator, whereas the polar coronal holes shrink and even disappear around the maxima, with streamers extending to all latitudes. As such, they serve as a basis to reconstruct the large-scale solar magnetic field and the hence that of the global solar wind (*e.g.*, Loucif and Koutchmy, 1989; Marsch, 2006; Lockwood and Owens, 2014; Hathaway, 2015; Owens *et al.*, 2017).

Both the MM and DM occurred long before the use of artificial coronagraphs which can reveal the coronal structure by blocking the bright solar disc. Such structures, however, can be revealed during total solar eclipses, when the Moon entirely hides the Sun and shut out most of its brightness. On such occasions, the brightness of the coronal streamers is visually captured (Eddy, 1976; Woo, 2019) and their extent provides valuable insight on the large-scale solar magnetic field (Owens *et al.*, 2017). The visual corona, as in unpolarised light, is a mixture of electron-scattered K-corona and dust-scattered F-corona. As such, extension of the K-corona is constrained by the structured solar magnetic field but F-corona appears structureless, free from such constraints.

Therefore, the coronal structure of the MM has attracted much scientific interest. Contemporary eclipse records have been intensively investigated and have shown the halo-shaped corona without significant streamer structure (Eddy, 1976; Riley *et al.*, 2015). Eddy (1976) speculated about a total loss of the solar magnetic field during the MM. Conversely, the continuation of solar cycles have been inferred from sunspot records and cosmogenic isotopes (Beer *et al.*, 1990, 1998; Usoskin *et al.*, 2001, 2015; Cliver and Ling, 2011; Lockwood *et al.*, 2011; Owens *et al.*, 2014; Vaquero *et al.*, 2015) and a report of a solar spicule or prominence during the 1706 eclipse (Foukal and Eddy, 2007), show that the large-scale solar magnetic field survived, even if its magnitude was greatly diminished (Cliver and Ling, 2011; Riley *et al.*, 2015; Hayakawa *et al.*, 2020).

In this context, the coronal structure in the DM is also of significant interest. However, eclipse reports in this period (*c.*, 1797 – 1827) have yet to be analysed with a view to understanding the large-scale solar magnetic field. Fortunately, this interval hosted significant developments in scientific understanding for the solar corona, when José Joaquín de Ferrer (1809) recorded the total eclipse on 1806 June 16. It was the extended nature of the glow around the eclipsed Sun that made the previously hypothesised association with an extended lunar atmosphere highly unlikely. From the work of Ferrer the name "corona" was established, as was the fact that it was part of the Sun (Vaquero and Vázquez, 2009). Moreover, de Ferrer was not a lone observer. Simeon de Witt (1809) also observed this eclipse and cited another graphical record. Situated in the midst of the DM, these records provide valuable visual evidence for the large-scale solar magnetic field. Therefore, we have conducted investigations on the eclipse records at that time, evaluated the reported coronal extents, and compare them with contemporary observations of sunspot number, as well as modelled reconstructions of the open solar flux, heliospheric modulation potential, and solar wind speed as a function of latitude and time.

## 2. Observations

The total eclipse on 1806 June 16 started from the coast of California, came across the central United States and the northern Atlantic Ocean, and ended in the Western Africa. Figure 1 shows its totality path, assuming the $\Delta T$ (difference of the terrestrial time and universal time) as 16.3 seconds (Stephenson *et al.*, 2016). As shown here, New England was favourably situated in this totality path and two notable eclipse drawings were recorded for this eclipse (see Figure 2).

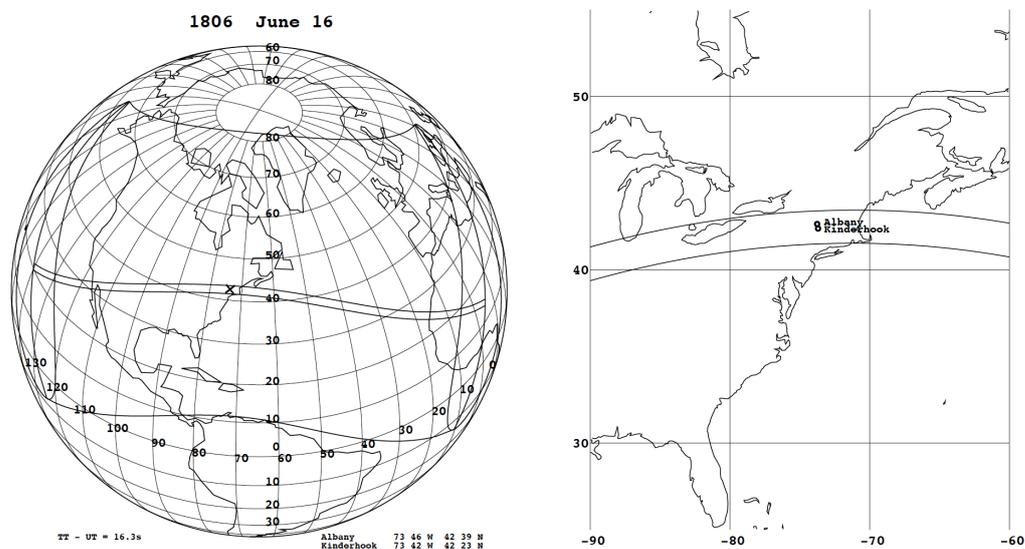

Figure 1: Totality path of the total eclipse on 1806 June 16, assuming the $\Delta T$ = 16.3 second (Stephenson *et al*., 2016) and its enlargement in the Eastern Coast of the United States. Albany and Kinderhook are marked in these maps.

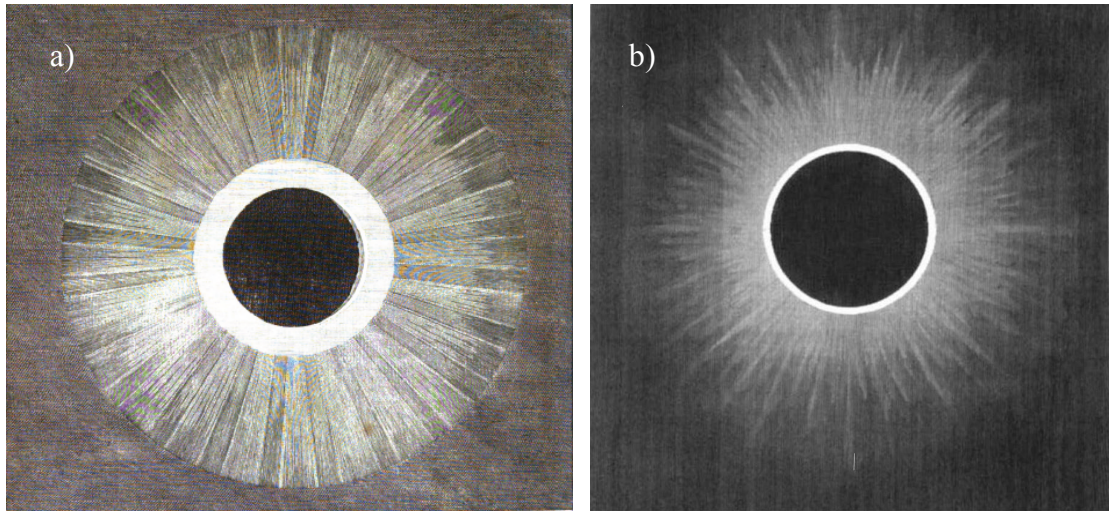

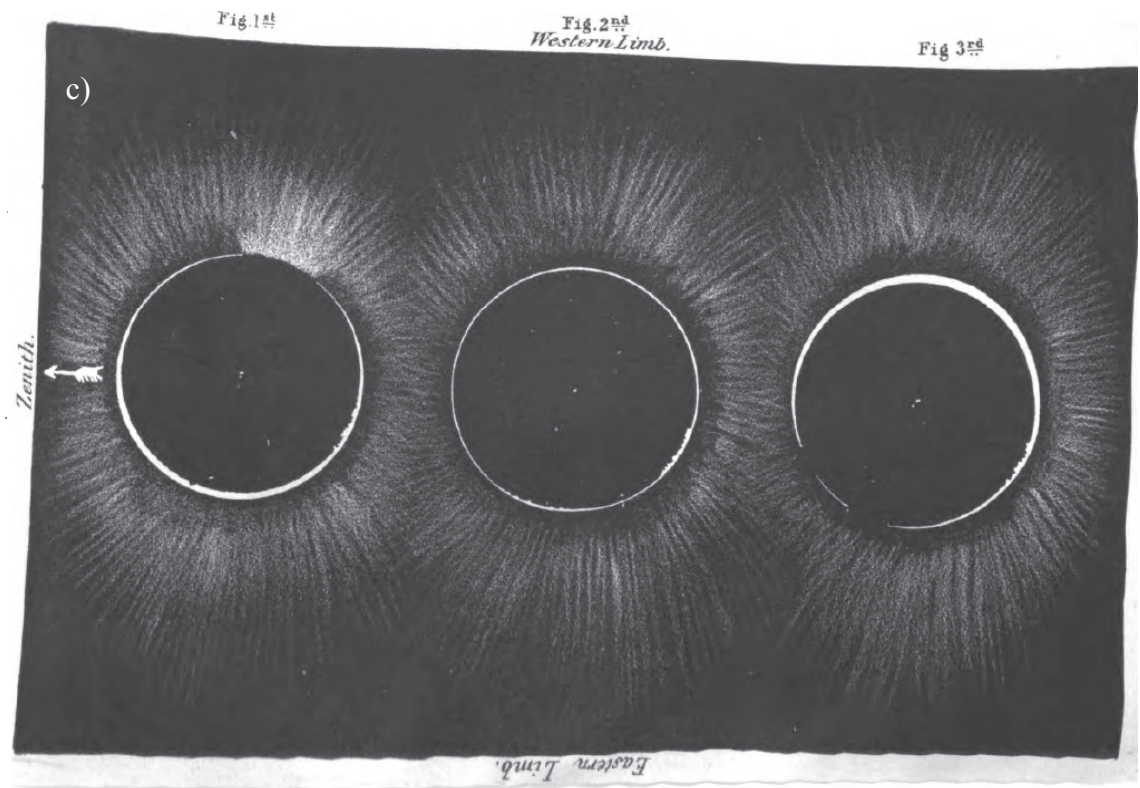

Figure 2: Total eclipse drawings on 1806 June 16; (a) Don José Joaquín de Ferrer's eclipse drawing reproduced from de Ferrier (1809, Plate VI, Figure 1); (b) and (c) Ezra Ames's eclipse drawings reproduced from de Witt (1852, Plate 3).

The first drawing is an original drawing of Don José Joaquín de Ferrer at Kinderhook (N42°23′, W73°42′. See Figure 2a), which has been often mentioned in the scientific literature (Todd, 1894; Vaquero and Vázquez, 2009). The drawing slightly emphasises the eclipsed Sun more than its deformed reproduction in Todd (1894, p. 115), which has been more often cited than the original version. De Ferrer used an achromatic telescope, a circle for reflection, an Arnold chronometer, and a darkened glass (De Ferrer, 1809, pp. 265 – 266). He described the eclipse thus; "the disk had round it a ring or illuminated atmosphere, which was of a pearl colour, and projected 6′ from the limb, the diameter of the ring was estimated at 45′. ... From the extremity of the ring, many luminous rays were projected to more than 3 degrees distance. The lunar disk was ill defined, very dark, forming a contrast with the luminous corona; with the telescope I distinguished some very slender columns of smoke, which issued from the western part of the moon. The ring appeared concentric with the sun, but the greatest light was; in the very edge of the moon, and terminated confusedly at 6′ distance. [At] 11:00, [I] observed the appearance of a ribbon or border, similar to a very white cloud, concentric with the sun, and which appeared to me to belong to its atmosphere, 90° to the left of the moon". (De Ferrer, 1809, pp. 266 – 267).

He emphasised the luminous ring around the eclipsed Sun: "Fig. 1 in Plate VI [NB our Figure 2a], represents the total eclipse, I shall only remark, that the luminous ring round the moon, is exactly as it appeared in the middle of the eclipse, the illumination which is seen in the lunar disk, preceded 6″ 8 the appearance of the first rays of the sun" (De Ferrer, 1809a, p. 274). "It has appeared to me, that the cause of the illumination of the moon, as noticed above, is the irradiation of the solar disk, and this observation may serve to give an idea of the extension of the luminous corona of the sun" (De Ferrer, 1809, p. 275).

This eclipse was also observed at Albany (N42°38′42″, W73°46′), where Ezra Ames painted and Simeon de Witt recorded its detail (Worth, 1866, p. 41). Ezra Ames was "an eminent portrait painter", as described by de Witt (1809, p. 300). His drawing was attached to de Witt (1809) and deposited in the Hall of the American Philosophical Society. Later on, his drawing has been involved in de Witt (1852, Plate 3) with a sequence of drawings, as shown in Figures 2b and 2c.

## 3. Results

These diagrams look consistent with each other, showing a brighter inner ring and the outer luminous rays or streamers all around the eclipsed Sun. Indeed, de Witt (1809, p. 300) emphasised its similarity with de Ferrer's drawing at Kinderhook. Observing from the same town, de Witt (1809) described his observations as: "The edge of the moon was strongly illuminated, and had the brilliancy of polished silver. No common colours could express this; I therefore directed it to be attempted as you will see, by a raised silvered rim, which in a proper light, produces tolerably well, the intended effect" (De Witt, 1809, p. 300); and "The luminous circle on the edge of the moon, as well as the rays which were darted from her, were remarkably pale, and had that bluish tint, which distinguishes the colour of quick-silver from a dead white" (De Witt, 1809, p. 301). De Witt's description of the colour is interesting as it fails to mention any red colour, which had been reported in the 1706 eclipse by Captain Stannyan and which reveals magnetic field in the chromosphere (see Foukal and Eddy, 2007).

The extent of the eclipse features is detailed in de Ferrer's report, along with their characteristics. The brighter inner ring reportedly extended ≈ 6′ with a colour of silver or pearl. The luminous rays had dimmer colour and reportedly extended from the inner ring with a distance of ≥ 3°. Although slightly stylised, their illustrations show the bright inner ring and the outer radiation (Figure 2). The breadth of the outer radiation is particularly notable. The inner and outer rings are probably best interpreted as lower solar atmosphere and the outer corona with streamers, respectively. Moreover, de Ferrer's description on "very slender columns of smoke, which issued from the western part of the moon" implies his observations on prominences or solar spicules (see *e.g.*, Beckers, 1968; Mackay *et al*., 2010).

The detailed reports on the visual extents of the inner ring and outer rays allow us to estimate their absolute extents. During the 1806 eclipse, the distances of the Sun and the Moon from Kinderhook were estimated as ≈ 1.0161892 au and ≈ 0.0023920 au with JPL DE430. Hence solar radius $R_\odot$ and lunar radius would span 15′44″ and 16′42″ in the sky, respectively. The maximal magnitude[1] at Kinderhook is calculated as ≈ 1.028, whereas this is calculated as ≈ 1.030 at the center-line near Kinderhook. Accordingly, the reported extent of the inner ring of ≈ 6′ from the lunar disk implies its absolute extent from the solar disk as ≈ 0.44 $R_\odot$, considering the difference of lunar and solar radii of 58″. Likewise, the reported extent of the outer rays of ≥ 3° from the limb of this inner ring implies its absolute extent from as ≥ 11.88 $R_\odot$.

## 4. Discussion

One of the striking common features of the eclipse reports is the coronal streamers all around the eclipsed Sun, captured both descriptively and graphically (Figure 2). This feature agrees well with the solar-maximum-type coronal structure (see *e.g.*, Figure 1 of Owens *et al.*, 2017). This supports the existence of a substantial the K-corona and hence large-scale solar magnetic field, even in the midst of the DM, unlike the records of the eclipse during the MM (Eddy, 1976; Riley *et al.*, 2015). On this basis, the DM could be considered in a similar state of the solar dynamo, only with reduced amplitude in comparison with the modern solar cycles, unlike the MM (*e.g.*, Riley *et al.*, 2015). This interpretation agrees with the existing discussion of the amplitude and duration of the solar cycles, as well as the sunspot distributions in the DM (Hayakawa *et al.*, 2020), in comparison with those of the MM (Eddy, 1976; Ribes and Nesme-Ribes, 1993; Usoskin *et al.*, 2015).

As shown in Figure 3, this eclipse occurred in the declining phase of SC 5, which peaked in 1805 February in smoothed monthly mean (Hathaway, 2015) of the international sunspot number (Clette *et al.*, 2014; Clette and Lefèvre, 2016; see Figure 3) as well as sunspot positions in Derfflinger's observations (Hayakawa *et al.*, 2020). This was also the case with frequency of reported mid-latitude aurorae in the European sector, on which basis John Dalton first noted the existence of this secular minimum and after whom it was subsequently named[2] (Dalton, 1834; Silverman, 1992). In fact, it is shown that auroral visibility generally moved poleward, both when compiling the existing auroral reports in the European sector, as well as those from Islands in the North-Eastern Atlantic Ocean (Lockwood and Barnard, 2015; Vázquez *et al.*, 2016).

---

[1] Here the magnitude of eclipse is defined by $(R_\odot + R_\leftmoon - d)/(2 R_\odot)$ where $R_\odot$ is the apparent angular radius of the Sun, $R_\leftmoon$ is the apparent angular radius of the moon, and $d$ is the apparent angular distance between the centers of the Sun and the Moon. In the case of partial solar eclipses the magnitude is equal to the fraction of the Sun's diameter obscured by the Moon. In the case of total solar eclipses the magnitude is equal to 1 at the instants of the beginning and end of the total solar eclipses and varies continuously with time.

Similar trends are found in centennial-scale reconstructions of solar activity based on a number of diverse sources. Cosmogenic isotopes, such as $^{14}$C and $^{10}$Be, can be used to estimate the time history of galactic cosmic ray (GCR) intensity reaching Earth, and thus the ability of the solar magnetic field to deflect GCRs (*e.g.*, Beer et al., 2012; Usoskin, 2017). This shielding ability is quantified by the heliospheric modulation potential (HMP). The shielding is actually caused by scattering of the GCRs by irregularities in the heliospheric field, but their net effect is well quantified by the open solar flux (OSF), the total solar magnetic flux that leaves to top of the solar atmosphere and fills the heliosphere and so acts as a barrier to GCRs. The faster deposition time of the $^{10}$Be cosmogenic isotope, and the fact that is not subsequently exchanged between different reservoirs, means that solar activity can potentially be resolved at annual timescales. However, a number of caveats apply in the interpretation of these data. The signal-to-noise in the $^{10}$Be records, coupled with the complexity of converting $^{10}$Be concentration into a measure solar magnetism means that at annual resolution the reconstructions contain uncertainties of the order ± 2 years in timing and around 25% in magnitude (Owens *et al.*, 2016b). The red and blue lines in the second panel of Figure 3 shows the annual HMP estimate from Muscheler *et al.* (2016) and decadal HMP estimate from Usoskin et al. (2014), while the black line shows the B (the near-Earth heliospheric magnetic field intensity, closely related to the OSF; see Figure 10 of Lockwood *et al.*, 2014) estimate from McCracken and Beer (2015), filtered in the same way as Owens *et al.* (2016b). While the same long-term trends are present in all cosmogenic estimates of solar activity, the annual reconstructions show less agreement about the timing and magnitude of individual cycles (see *e.g.*, Beer *et al.*, 1990; Berggren *et al.*, 2009).

---

[2] It is Sam M. Silverman who suggested this term during his discussion with Jack Eddy and George Siscoe (private communication with S. M. Silverman in 2020).

OSF and near-Earth heliospheric field, B, can also be estimated from sunspot records, by using assuming sunspots represent the source of new OSF and that OSF can be treated as a continuity equation (Solanki *et al*, 2000). This method gives very good agreement with geomagnetic reconstructions over the interval 1845-2013 (Owens *et al*., 2016a). Of course, there may be long-term drifts in the calibration of the sunspot record before this period (from changes in observing capability, intercalibration of different observers, *etc*.; see Clette *et al*., 2014; Clette and Lefèvre, 2016), which makes the independent estimates of cycle amplitude from $^{14}$C and $^{10}$Be very useful. However, the timing of sunspot cycles, and hence features in the subsequent OSF reconstruction, likely accommodate uncertainty of a few years for the epoch of DM (Adolphi and Muscheler, 2016).

The third panel of Figure 3 shows that the open solar flux (OSF) from the model constrained by the sunspot number did not peak until mid 1806, when this eclipse took place. Here, the OL12 model (Owens and Lockwood, 2012) has been applied to different sunspot series: Svalgaard and Schatten (2016), Lockwood *et al*. (2014), and SILSO V2 (Clette and Lefevre, 2016), and shown in red, black, and blue curves, respectively. The red and the black curves thus correspond to the "high" and "low" scenarios in Asvestari *et al*. (2017). These reconstructions are compared with the grey-shaded region which is the modelled OSF from Vieira *et al*. (2011), based on the SATIRE-T model applied to the group sunspot number of Hoyt and Schatten (1998; HS98). All these OSF reconstructions are unsigned flux. Here, the OSF from Vieira *et al*. (2011) shows slightly lower value in comparison with other curves with OL12 model, as the HS98 series used in Vieira *et al*. (2011) shows a larger trend netween 1800 and now than the other series used in OL12 model (Svalgaard and Schatten, 2016; Lockwood *et al*., 2014; SILSO V2), as shown in Figure 11 of Clette and Lefèvre (2014).

Further information about the expected structure of the corona and solar wind can be estimated by assuming new OSF is produced in the streamer belt, resulting in slow wind, which then gradually transitions into coronal hole flux, resulting in fast solar wind (Lockwood and Owens, 2014). The time constant for this transition is a free parameter determined by comparison with 40 years of photospheric magnetic field observations and models (see Owens *et al*., 2017). The resulting solar wind structure as a function of latitude and time is shown in the fourth panel of Figure 3. On this basis, the eclipse occurrence in mid 1806 occurs during an interval with slow wind at most latitudes, suggesting streamers should extend to most latitudes. This is broadly consistent with the eclipse images (Figure 2), which showed streamers all around the eclipsed Sun. As such, these two eclipse drawings in 1806 June confirm the validity of the existing models of Owens *et al*. (2017) within the DM in terms of their reconstructions of OSF phase and solar-wind speed as a function of latitude and time.

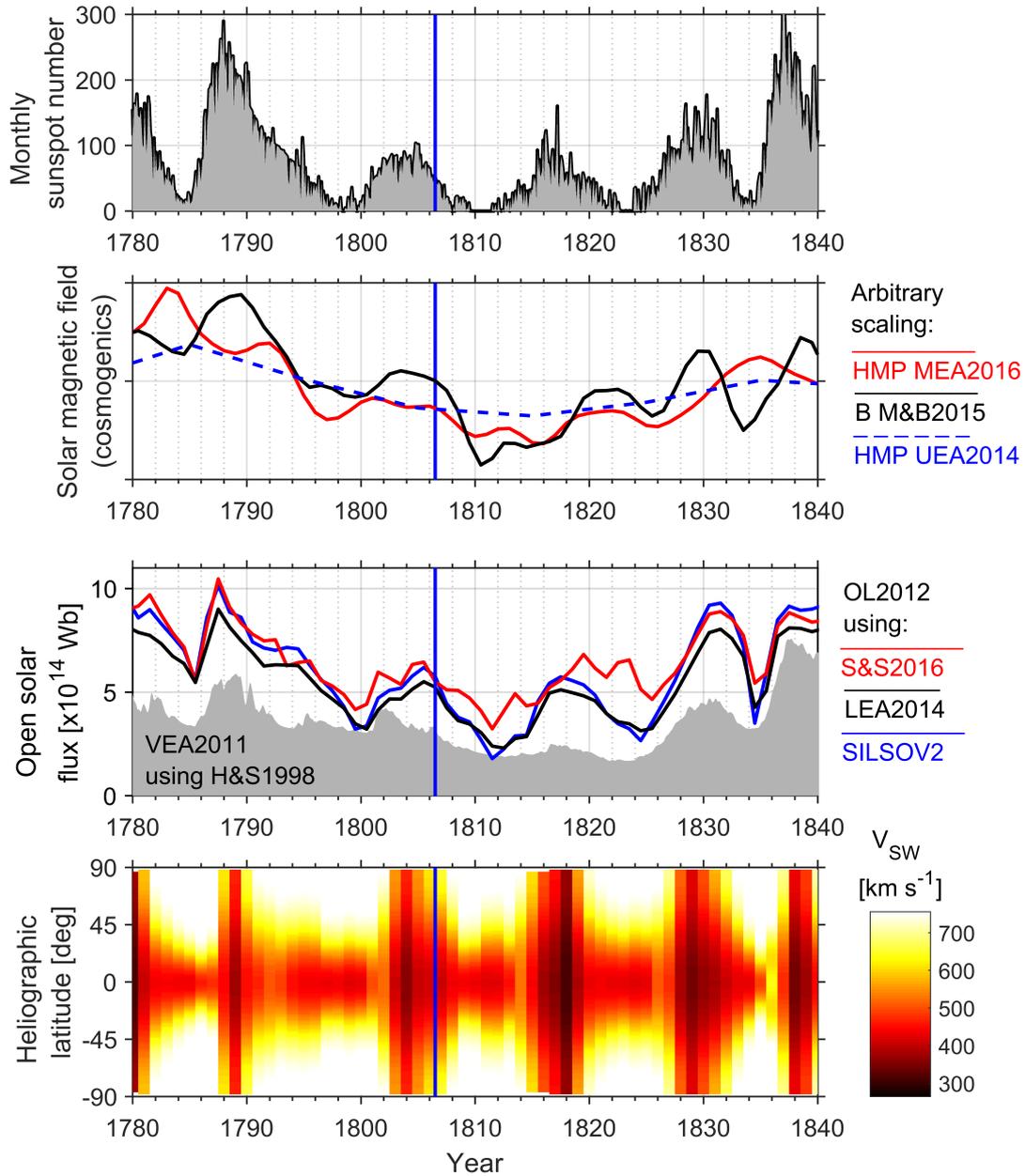

Figure 3: A summary of observed and modelled solar properties through the Dalton Minimum. The 1806 eclipse is shown as the blue vertical line. First panel: Monthly sunspot number (Clette and Lefèvre, 2016). Second panel: The coloured lines show estimates of solar activity, scaled for plotting purposes: HMF B from $^{10}$Be (McCracken and Beer, 2015; Owens *et al*., 2016b; black), annual (red) and decadal (blue) heliospheric modulation potential from $^{14}$C (Muscheler *et al*., 2016; Usoskin *et al*., 2014). Third panel: Reconstructed open solar flux based on the OL model (Owens and Lockwood, 2012), applied to different sunspot series: red = Svalgaard and Schatten (2016), black = Lockwood *et al*. (2014), and blue = SILSO V2 (Clette and Lefèvre, 2016). Thus the red and the black curves correspond to the "high" and "low" scenarios in Asvestari *et al*. (2017). The grey-shaded region is the modelled OSF from Vieira *et al*. (2011), based on HS98. Fourth panel: The reconstructed solar wind speed as a function of heliographic latitude and time (Owens *et al*., 2017).

## 4. Conclusion

In this article, we have examined the total eclipse drawings on 1806 June 16 and visually confirmed the activity phase of the solar magnetic field in the midst of the DM. Both of de Ferrer's and Ames's eclipse drawings showed corona with significant rays and streamers. On the basis of de Ferrer's report, we computed the extent of the brighter and the outer rays from the solar disk as $\geq 11.88\ R_\odot$, and $\approx 0.44\ R_\odot$, respectively. De Ferrer's report also implies presence of prominences or solar spicules. These details confirm the presence of the solar and heliospheric magnetic fields in the midst of the DM.

This marks a significant difference from the coronal structure during the MM, when streamers were apparently missing or at least not bright enough to be visible and the corona was recorded without significant structure. This contrast visually shows significant difference of the DM with the MM in terms of their background state of the solar dynamo, and robustly supports the existing discussions on the difference of the DM and MM on the basis of their sunspot positions and amplitude and duration of their solar cycles (Usoskin *et al*., 2015; Hayakawa *et al*., 2020). This comparison disproves postulates that the MM was no more than an extended version of the DM such that both are similar minima of the quasi-regular Gleissberg cycle (Zolotova and Ponyavin, 2015): the same conclusion was reached by Usoskin *et al.* (2015) looking at a variety of other historic and paleo- datasets

Moreover, comparison these eclipse drawings is broadly consistent with the modelled reconstruction on the cycle phase of OSF and on that on the solar wind speed as a function of latitude and time. The OSF peaked around this eclipse and the slow solar wind extended to most latitudes, suggesting streamers should also extend to most latitudes. This coincidence confirms the validity of the existing model of Owens *et al.* (2017) even in the midst of the DM.

**Acknowledgement**

We thank Ken'ichi Fujimori for his advice on the coronal visibility during the total eclipses in 1991 and 2009, WDC SILSO at Royal Observatory of Belgium for providing international sunspot number and its regular maintenance, Joe DiLullo and other archivists in American Philosophical Society Archives for their advices on the eclipse reports, Sam M. Silverman for letting us know the background history on how the Dalton Minimum was named, and Raimund Muscheler for providing the background data in Muscheler *et al.* (2016). We thank Ilya Usoskin for providing background data for Vieila et al. (2011) and Usoskin et al. (2014) and Shin Toriumi for his helpful discussions and suggestions. HH was part-funded by Young Leader Cultivation (YLC) program of Nagoya University, the 2019 Collaborative Research Grants for YLC (grant # YLC2019A02), the Unit of Synergetic Studies for Space of Kyoto University, BroadBand Tower, and JSPS grant-in-aids (JP15H05812 and JP15H05816). MO was part-funded by Science and Technology Facilities Council (STFC) grant number ST/R000921/1.